\documentclass[final]{svjour3}
\usepackage{graphicx}
\usepackage{rotating}
\usepackage{amssymb}
\usepackage{mathptmx}
\usepackage[numbers]{natbib}
\makeatletter
\journalname{Journal of Low Temperature Physics}

\bibpunct{}{}{,}{s}{}{,}
\newcommand*\Laplace{\mathop{}\!\mathbin\bigtriangleup}
\newcommand*\conj[1]{\overline{#1}}

\begin{document}

\newcommand{\hdblarrow}{H\makebox[0.9ex][l]{$\downdownarrows$}-}
\title{A Composite Phononic Crystal Design for Quasiparticle Lifetime Enhancement in Kinetic Inductance Detectors}

\author{T.A. Puurtinen$^\textbf{1}$ \and K. Rostem$^\textbf{2}$ 
\and P.J. de Visser$^\textbf{3}$ \and I.J. Maasilta$^\textbf{1}$}

\institute{$^\textbf{1}$ Nanoscience Center, Department of Physics, University of Jyv\"askyl\"a, FI-40014, Finland\\
$^\textbf{2}$ NASA Goddard Space Flight Center, 8800 Greenbelt Road, Greenbelt, Maryland 20771, USA\\
$^\textbf{3}$ SRON, Netherlands Institute for Space Research, Sorbonnelaan 2, 3584 CA, Utrecht, The Netherlands\\
\email{maasilta@jyu.fi}}

\maketitle

\begin{abstract}

A nanoscale phononic crystal filter (reflector) is designed for a kinetic inductance detector where the reflection band is matched to the quasiparticle recombination phonons with the aim to increase quasiparticle lifetime in the superconducting resonator. The inductor is enclosed by a 1 $\mu$m wide phononic crystal membrane section with two simple hole patterns that each contain a partial spectral gap for various high frequency phonon modes. The phononic crystal is narrow enough for low frequency thermal phonons to propagate unimpeded. With 3D phonon scattering simulations over a 40 dB attenuation in transmitted power is found for the crystal, which was previously estimated to give a lifetime enhancement of nearly two orders of magnitude. 


\end{abstract}

\section{Introduction}

Phononic crystals (PnCs) are periodic structures designed to modify properties of elastic and acoustic waves by the Bragg scattering mechanism or by localized resonances. \cite{kushwaha} Applications have been proposed in many engineering and physics areas, including vibration and noise absorption, acoustic cloaking and metamaterials research. \cite{craster} 
Recently, with the ability to fabricate microscopic phononic crystals,  thermal applications have also appeared benefiting from the wave interference of phonons at thermal wavelengths. This could be utilized  for instance in improving thermoelectric devices by reduction of thermal conductivity, guiding and focusing heat in nanostructures and using heat in phononic information processing. \cite{thermo, roman, info} At sub-Kelvin temperatures phononic crystals fabricated in a suspended membrane have been shown to reduce thermal conductance by several orders of magnitude \cite{natcom, yaolan},
which is key to improving sensitivity 
in many low temperature bolometric detector technologies. 

Kinetic inductance detectors (KID), on the other hand, are extremely sensitive non-bolometric photon detectors based on superconducting microwave resonators. \cite{day} The operation principle of a KID is to break Cooper pairs into quasiparticles in a superconducting microwave resonator with signal photons, thus changing the impedance of the resonator, which can be measured with a coupled microwave readout line. Quasiparticles recombine quickly back into Cooper pairs by emitting phonons at frequencies $\geq 2\Delta/h$, corresponding to the energy gap of the superconducting material. The recombination phonons can break other Cooper pairs, but phonons escaping the system result in a lost signal limiting the responsivity of the detector. Simply removing material around the resonator does not help because a good thermal contact is desired so that low-frequency phonons due to dissipation of the microwave readout signal can escape. Recently Rostem et al. presented calculations for an effective recombination life time enhancement of quasiparticles in a phononic crystal isolated kinetic inductance detector (PhiKID). \cite{rostem} They showed how a phononic crystal notch filter with a $0.3\Delta/h$ wide reflection band and a 40~dB attenuation centered at $2\Delta/h$ of a BCS superconductor can increase the life time by two orders of magnitude. 

A number of publications in the field of ultrasound physics discuss the formation of bandgaps or scattering of elastic waves in plates with obstacles \cite{gorishnyy, pennec, laude}, which are useful starting points in finding suitable PnC geometries with full band gaps for the recombination phonons. However, it will be difficult to fabricate such PnCs for the most commonly used low-$T_\textrm{c}$ superconductors, such as aluminium, which has $2\Delta/h\sim 89$~GHz. For example, creating a matched reflection band for aluminium with a thru-hole pattern in a suspended silicon nitride membrane would steer the fabrication towards very thin membranes ($\sim 33$~nm) and the smallest features in a few nanometer length scale. \cite{natcom} Even for thin-film hafnium, which has $2\Delta/h \sim 32$~GHz, a matching PnC design with a thru-hole pattern would require a $95$~nm thick membrane with a distance between holes only 5~nm. 

In this paper, we propose a simulation based design for a composite PnC consisting of two simple hole patterns with minimum feature size 50~nm, and a relatively low combined width of 1~$\mu$m. Each of the patterns possess partial spectral gaps for various modes, which effectively create a full reflection band for high frequency athermal recombination phonons, while thermal low-frequency phonons pass through the structure unimpeded. Interestingly, a PnC will also reject athermal phonons released during the energy downconversion of quasiparticles excited to energies significantly greater than $\Delta$. The downconversion process releases phonons with at the scale of the Debye frequency of the metal. The loss of these so-called hot phonons from the superconductor to the bath is known to limit the performance of many cryogenic detectors, including STJs \cite{martin},  TESs \cite{kozorezov}, and KIDs \cite{guo}. A PnC can effectively recycle hot phonons through a purely geometric mechanism where the membrane crossectional area is reduced, at least due to the lack of material that is etched away. At the same time, due to the scale of the PnC pattern, thermal phonons with frequencies below 10~GHz have unity transmission. Thus, a PnC may be effective at improving the performance of KIDs that utilize Al, or other high temperature superconducting metals, to detect visible and near-infrared photons.

\section{Scattering Simulations}

\begin{figure}
\begin{center}
\includegraphics[width=0.8\linewidth, keepaspectratio]{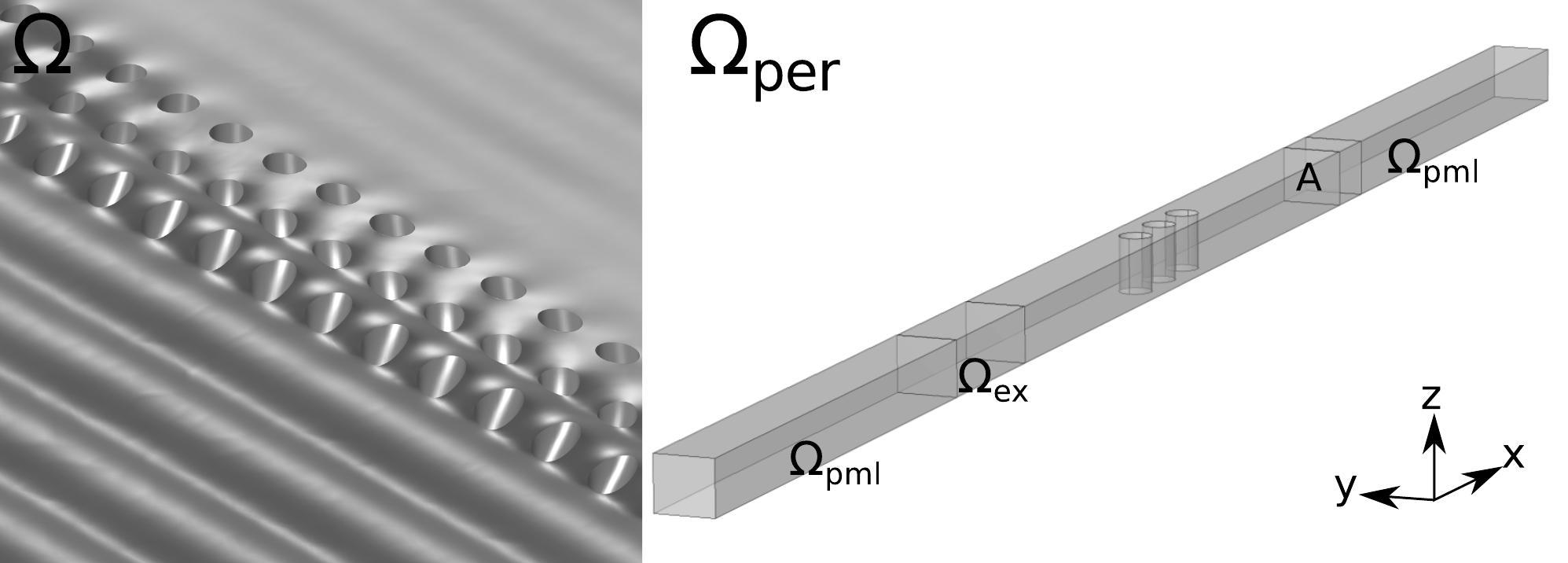}
\caption{(color online) Left: 3D simulation of Lamb wave scattering in an infinite phononic crystal membrane. Right: Truncated periodic subdomain $\Omega_\textrm{per}$ of the membrane. An incident phonon is excited at the $\Omega_{\textrm{ex}}$ region, and the scattered field from the crystal is absorbed in perfectly matched layers $\Omega_\textrm{pml}$.}
\label{fig1}
\end{center}
\end{figure}

Phonon propagation and scattering in PnC membranes at sub-Kelvin temperatures is described with a 3D continuum linear elasticity theory. \cite{graff} For smooth membranes, i.e. where the surface roughness is small compared to the phonon wavelength, phonon mean free paths have been measured to exceed 100 $\mu$m, \cite{Hoevers} thus phonons can be treated as ballistic waves that can only scatter from the PnC structure. The membrane material can be assumed to be homogeneous and it is characterized by a constant density $\rho$ and elasticity tensor $C_{ijkl}$. For an isotropic membrane material, tensor $C_{ijkl}$ can be reduced to two independent parameters, often denoted by the Lame parameters $\lambda$ and $\mu$. The displacement field $U = (u,v,w)$ in such system obeys
\begin{equation}
\label{lame}
\lambda\Laplace U + (\lambda + \mu)\nabla\cdot \textrm{div}\, U + \rho F = \rho\frac{\partial^2 U}{\partial t^2}, \qquad \textrm{in } \Omega
\end{equation}
where $\Omega$ is the domain describing the infinite PnC membrane geometry (see. Fig.~\ref{fig1} left) and $F = F(x,y,z,t)$ is the driving force applied in the domain. On the domain boundaries (the top and bottom surfaces, and hole boundaries) we apply the stress-free boundary condition $\hat{n}\cdot\sigma(U) = 0$. Here $\sigma(\cdot) = (\sigma_{ij}(\cdot))$, $i,j = x,y,z$ is the stress tensor and $\hat{n}$ is the surface normal vector. Assuming time harmonic force $F$ and solution $U$, the derivative $\partial^2/\partial t^2$ can be replaced by $-\omega^2$, where $\omega$ is the angular frequency of the force. \cite{graff}

The time harmonic problem of Eq. (\ref{lame}) is solved in the frequency domain using the finite element method and COMSOL Multiphysics v5.3 software. When the incident wave strikes the PnC from a 90 degree angle (Fig. ~\ref{fig1} left), the scattered field is periodic in $\Omega$ in the $y$-direction, and thus 
the infinite domain $\Omega$ can be truncated as shown in Fig.~\ref{fig1} (right).
Periodic boundary conditions for the solution $U$ are set on the $x$-$z$ boundaries so that unphysical coupling or conversion of the modes does not occur there. Finally, perfectly matched layers $\Omega_\textrm{pml}$ are created at the ends to absorb the scattered phonons. We remark that it is in principle possible to simulate the system in a full scale 3D model without periodicity allowing variable incident angles, but this would greatly increase the computational cost.

To study reflection and mode conversion properties of PnCs we aim to run the scattering simulations with pure shear and Lamb modes \cite{graff}, which are the solutions for Eq.~(\ref{lame}) in the regions far away from the PnC. However, a typical problem with excitation of pure modes is that the scattered field re-scatters from the emitter and thus interferes with the original incident phonon. Neglecting the interference can lead to incorrect evaluation of transmission and reflection coefficients. We solved this problem by taking advantage of the biorthogonality property of the modes in the force term $F$ applied in the 3D subdomain $\Omega_\textrm{ex}$. \cite{gregory,hirose}
If the solution $U$ on $\Omega_\textrm{ex}$ is expressed as $U = U_\textrm{inc}^+ + U_\textrm{scat}$, where
$U_\textrm{scat} = \sum_{j}\alpha_jU_j^-$
and plus (minus) denotes waves propagating in the positive (negative) $x$-direction, then by the biorthogonality property
\begin{equation}
\label{biortho}
\int_{\Omega_\textrm{ex}} U_j^-\cdot\conj{\sigma_x(U_\textrm{inc}^+)}\,\,\textrm{d}V - \int_{\Omega_\textrm{ex}} \conj{U_\textrm{inc}^+}\cdot \sigma_x(U_j^-)\,\, \textrm{d}V = 0
\end{equation}
for all mode indices $j$. Here $\sigma_x(\cdot) := (\sigma_{xx}(\cdot),\sigma_{xy}(\cdot),\sigma_{xz}(\cdot))^\textrm{T}$ is a vector of the stress tensor components in the $x$-direction, and the overline denotes the complex conjugate. In the weak formulation of the elasticity equation (\ref{lame}) (see e.g. \cite{hughes}), the domain force term appears in the same form as the first term of the biorthogonality property (\ref{biortho}), if the force term is set to $F = \conj{\sigma_x(U_\textrm{inc}^+)}$ and the functions $U_j^-$ are replaced by the finite element basis functions. We then add another integral term in the weak formulation corresponding to the second term of the biorthogonality property, which together with the first term guarantees that all scattered wave components $U_j^-$ in the solution $U$ result to zero in these integrals and thus cannot backscatter from the subdomain $\Omega_\textrm{ex}$. Phonon modes $U_j^\pm$ in the membrane can be calculated either numerically, or semi-analytically by solving the Rayleigh-Lamb dispersion equations \cite{graff}. After normalization, functions $U_j^\pm$ are inserted into the weak formulation integrals for a numerical solution of the scattering problem.

Similarly, the phonon transmission power can be evaluated by the biorthogonality relation. \cite{hirose} In fact, the mechanical power of the elastic field $U$ carried by the mode $U_j^+$ through a boundary $A$ can be expressed as
\begin{equation}
P_j = \Re\left[\frac{i\omega}{4}\int_{A}\left(U\cdot\conj{\sigma_x(U_j^+)} - \conj{U_j^+}\cdot \sigma_x(U)\right) \,\,\textrm{d}A\right]
\end{equation}
and the total power is then the sum of single mode contributions $P_j$.

\section{Results}

To find a practical design for the PnC membrane with good reflective properties in the 32~GHz region (corresponding to the energy gap of thin-film hafnium), we numerically calculated phonon dispersion relations for several simple infinite cylindrical hole patterns in crystalline silicon membranes varying the lattice constant, hole filling fraction, membrane thickness and the lattice geometry. The designs with a full spectral gap at 32~GHz found this way were not feasible for fabrication with current lithography processes because of too small feature sizes going below 5~nm length scale. A full spectral gap, however, is not mandatory, because a similar effect can be achieved by combining several simpler structures in series each possessing a partial spectral gap for some Lamb modes.
With the scattering simulation we can then estimate the effect of the composite structure for each phonon mode.

\begin{figure}
\begin{center}
\includegraphics[width=0.8\linewidth, keepaspectratio]{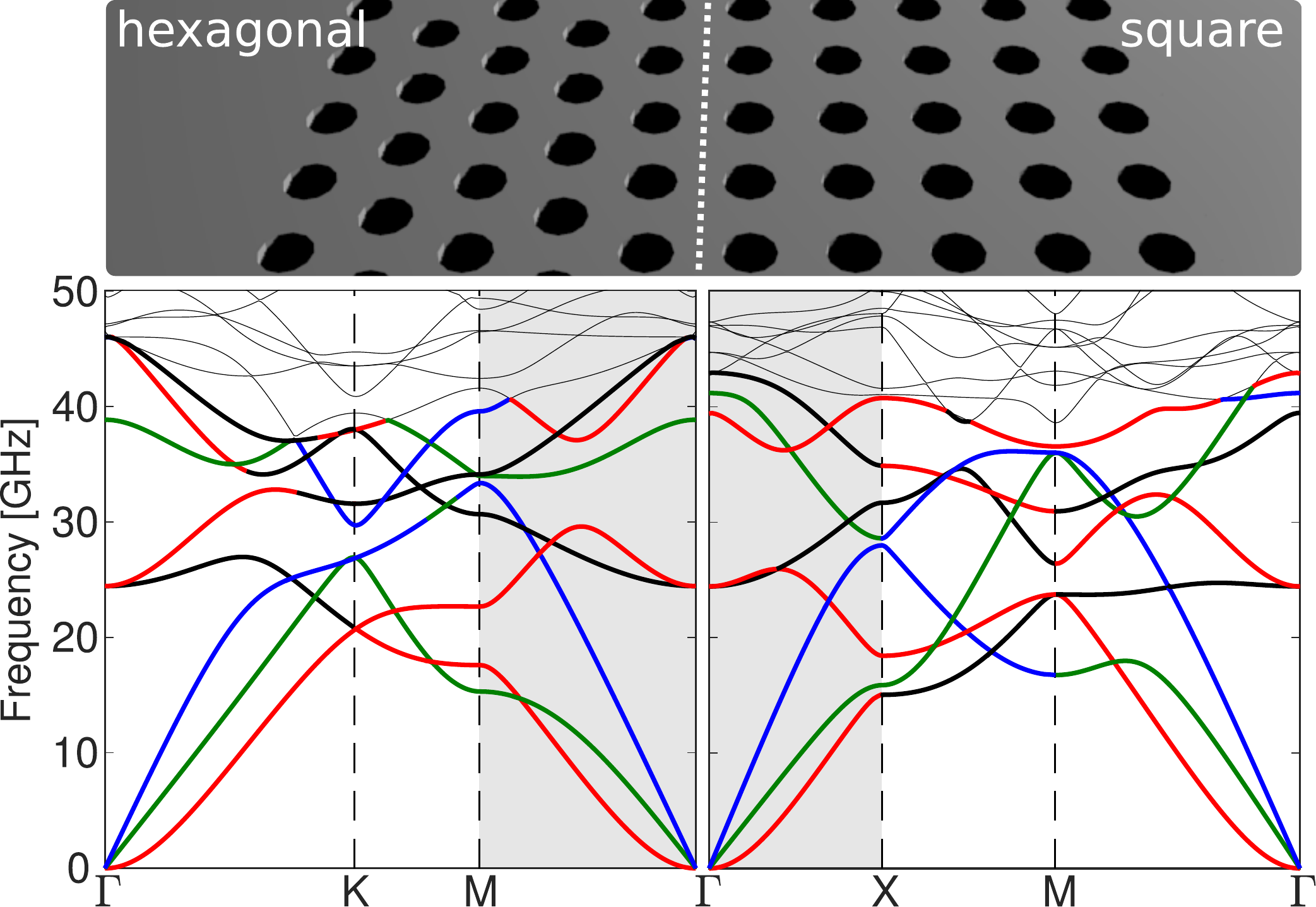}
\caption{(color online) Top: A composite phononic crystal structure containing cylindrical holes in square and hexagonal lattice configurations. Left: Spectrum of the hexagonal lattice. Right: Spectrum of the square lattice. For color coding, see text.}
\label{fig2}
\end{center}
\end{figure}

The most practical geometry found, shown in Fig.~\ref{fig2},
is a composite thru-hole PnC design, which functions as a phononic multi-pole filter. \cite{karwan1D} The geometry contains cylindrical holes in hexagonal and square lattices in a thin silicon membrane with thickness 100~nm. The lattice constant for both lattices is 110~nm and the hole diameter is 60~nm. The square lattice is 5 cells and the hexagonal lattice 5 cells wide making the composite crystal only about 1 $\mu$m wide with the smallest feature (neck width) 50~nm. The corresponding dispersion spectra are also shown for each of the components. This geometry can be fabricated with a single exposure etch process.

It is immediately seen from the dispersion relations that at the low-frequency (sub-$10$~GHz) range the spectra resemble a full membrane spectrum, \cite{graff} thus the PnC should have minimal effect on thermal conductance in the membrane. In addition, neither of the lattices seem to have a full spectral gap at 32~GHz. However, further analysis of the modes reveals partial spectral gaps, that can be utilized for the same effect. We remark that pure Lamb modes do not exist in the PnCs, but in these lattices (anti-)symmetrically polarized incident waves are likely to remain (anti-)symmetric also in the PnC. In other words, if there are no propagating antisymmetric out-of-plane modes in the PnC, then it can be expected that incident antisymmetric out-of-plane Lamb modes are reflected from the PnC. To locate the partial spectral gaps in Fig.~\ref{fig2}, we classified the modes based on a component analysis of the fields $U$. In Fig.~\ref{fig2} red corresponds to the (mainly) antisymmetric out-of-plane mode, green is the in-plane shear, blue is the symmetric out-of-plane mode and black is the second order antisymmetric out-of-plane mode. The grey highlighted area shows the direction perpendicular to the PnC, which can be simulated with the scattering model. Following the colored curves, partial spectral gaps can be identified for each of the mode types in this region.

\begin{figure}
\begin{center}
\includegraphics[width=0.8\linewidth, keepaspectratio]{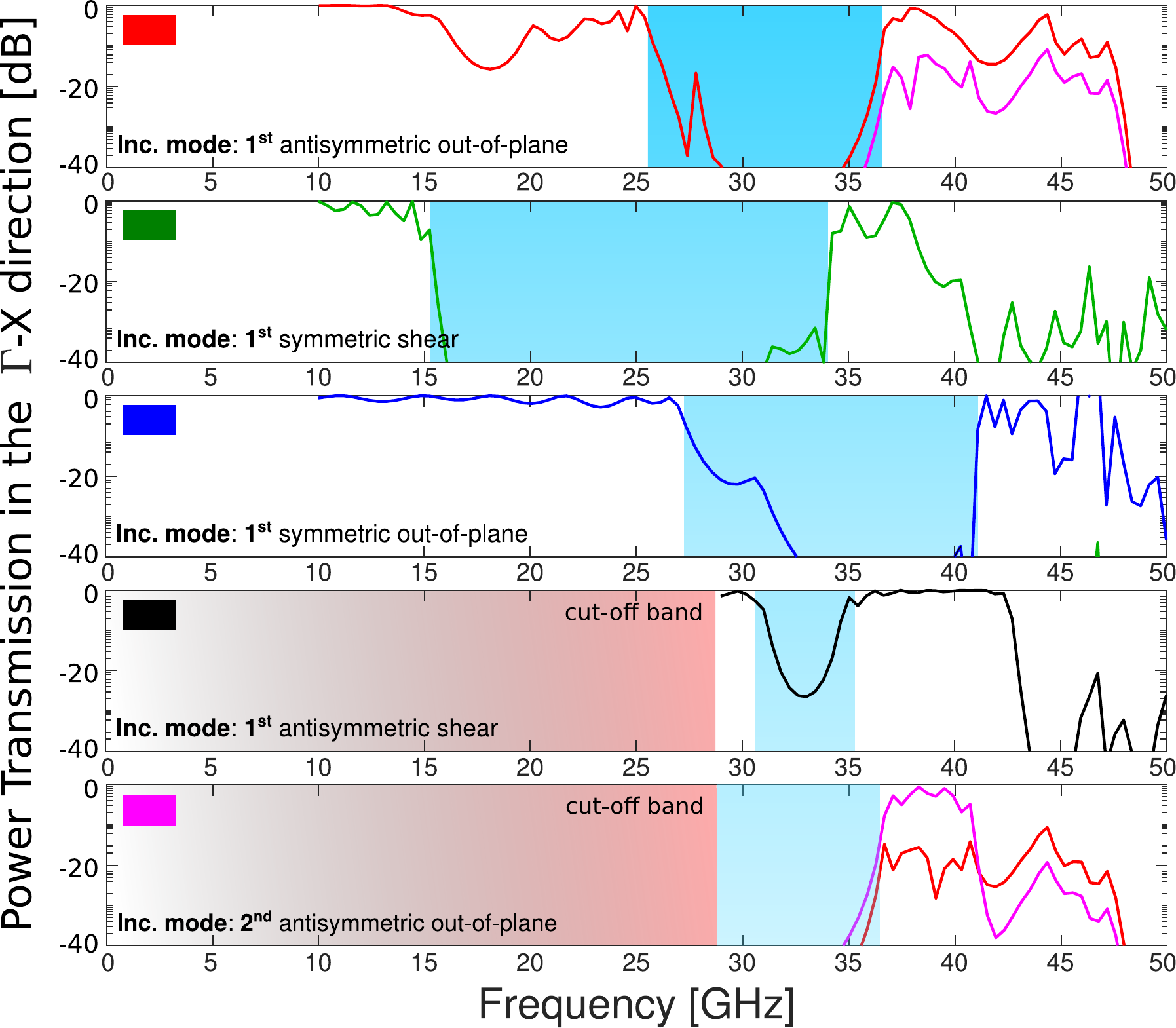}
\caption{(color online) Transmitted phonon power through the composite phononic crystal of Fig. ~\ref{fig2} with normal incidence. Colored curves in the plots reveal the amount of coupling: e.g. in the first plot, the 1st antisymmetric out-of-plane mode (red) couples with the 2nd order mode (magenta) at $f > 35$ GHz. For all modes there exist a significant reflection band at 32 GHz, while at the low frequency limit power transmission approaches unity leading to good thermal conductance.}
\label{fig3}
\end{center}
\end{figure}

Finally, the scattering simulation is used to calculate the attenuation level of the composite structure. In Fig.~\ref{fig3} the transmitted power of each of the propagating Lamb and horizontal shear modes are shown around the 32~GHz frequency region.
Colors of the various curves denote different mode polarizations, and the incident mode color is shown at the top-left corner. For all modes a fairly wide reflection band is created at 32 GHz, which is highlighted by blue.  Cut-off bands for the second order modes are highlighted by red, which means that these modes do not exist at low frequency range. Incident modes in most cases forward-scatter to the same mode, but a small amount of coupling can be identified for the first and the second order antisymmetric out-of-plane modes as revealed by the secondary curves in those plots.
We remark that changing the incident angle slighty will not cause significant qualitative changes to the transmitted power in Fig.~\ref{fig3}, and similar attenuation levels and band locations are likely to exist for a range of incident angles. 

To obtain an estimate of the effective quasiparticle lifetime enhancement for each phonon mode, for each of curves in Fig.~\ref{fig3} reflection band widths and locations can be inserted into the equations in Rostem et al. \cite{rostem} A single effective reflection band location and width parameters for the total power cannot be formulated from this data, because the real emitted phonon angular distribution and population in various mode types is not calculated in this study. In a PhiKID detector, however, the antisymmetric modes are likely to be dominant, because the metal thin-film emitter is located only on the top-side of the suspended membrane, thus making the source geometry antisymmetric relative to the membrane. For an estimate of the total effective lifetime enhancement, the first antisymmetric mode can be chosen.

\section{Conclusion}

Nanoscale phononic crystals (PnCs) appear as promising functional components in many low temperature detector technologies. Kinetic inductance detectors have been suggested to benefit from notch filter type PnCs when the filter is tuned to the quasiparticle recombination phonon frequencies of the KID superconducting resonator. However, for many superconductor materials fabricating such a filter is challenging even with state-of-the-art nanofabrication techniques, because of extremely small features required for the PnC. Using 3D elastic wave scattering simulations we demonstrated how a composite PnC assembled from two simple hole patterns can reach at least 40 dB power attenuation level for the recombination phonons incident from the perpendicular direction, which has been estimated to result in effective quasiparticle lifetime enhancement of two orders of magnitude. The smallest feature in this design was only 50 nm and the structure was only 1 $\mu$m wide making it relatively simple to fabricate with modern electron beam lithography techniques. Based on these results, a phononic crystal isolated kinetic inductance detector could be realized in the near future.  

\begin{acknowledgements}
This study was supported by the Academy of Finland Project Number 298667. K. Rostem gratefully acknowledges financial support from a NASA Astrophysics Research and Analysis grant (NNX17AH83G).
P. J. de Visser was financially supported by the Netherlands Organisation for Scientific Research NWO (Veni Grant 639.041.750).
\end{acknowledgements}



\begin{thebibliography}{99}

\bibitem{kushwaha}
M.S. Kushwaha, P. Halevi, L. Dobrzynski, and B. Djafari-Rouhani, {\it Phys. Rev. Lett.} \textbf{71}, 2022, (1993), 
DOI:10.1103/PhysRevLett.71.2022.

\bibitem{craster}
R.V. Craster, and S. Guenneau, (eds.) Acoustic Metamaterials, vol. 166 of Springer Series in Material Science, Springer, (2013).

\bibitem{thermo}
A.J. Minnich, M.S. Dresselhaus, Z.F. Ren, and G. Chen, {\it Energ. Environ. Sci.} \textbf{2}, 466-479, (2009),
DOI:10.1039/B822664B.

\bibitem{roman}
R. Anufriev, A. Ramiere, J. Maire, and M. Nomura, {\it Nat. Commun.} \textbf{8}, 15505, (2017),
DOI:10.1038/ncomms15505.

\bibitem{info}
N. Li, J. Ren, L. Wang, P. H\"anggi, and B. Li, {\it Rev. Mod. Phys.} \textbf{84}, 1045, (2012),
DOI:10.1103/RevModPhys.84.1045.

\bibitem{natcom}
N. Zen, T.A. Puurtinen, T.J. Isotalo, S. Chaudhuri, and I.J. Maasilta,
{\it Nat. Commun.} \textbf{5}, 3435, (2014), 
DOI:10.1038/ncomms4435.

\bibitem{yaolan}
Y. Tian, T.A. Puurtinen, Z. Geng, and I.J. Maasilta, {\it Phys. Rev. Appl.} \textbf{12}, 014008, (2019),
DOI:10.1103/PhysRevApplied.12.014008.

\bibitem{day}
P.K. Day, H.G. LeDuc, B.A. Martin, A. Vayonakis, and J. Zmuidzinas, {\it Nature} \textbf{425}, 817, (2003),
DOI:10.1038/nature02037.

\bibitem{rostem}
K. Rostem, P.J. de Visser, and E.J. Wollack,
{\it Phys. Rev. B} \textbf{98}, 014522, (2018), DOI:10.1103/PhysRevB.98.014522.

\bibitem{gorishnyy}
T. Gorishnyy, C.K. Ullal, M. Maldovan, G. Fytas, and E.L. Thomas,
{\it Phys. Rev. Lett.} \textbf{94}, 115501, (2005),
DOI:10.1103/PhysRevLett.94.115501.

\bibitem{pennec}
Y. Pennec, B. Djafari-Rouhani, H. Larabi, J.O. Vasseur, and A.C. Hladky-Hennion, {\it Phys. Rev. B} \textbf{78}, 104105, (2008),
DOI:10.1103/PhysRevB.78.104105.

\bibitem{laude}
R.P. Moiseyenko, S. Herbison, N.F. Declercq, and V. Laude, {\it J. Appl Phys.} \textbf{11}, 034907, (2012),
DOI: 10.1063/1.3682113.

\bibitem{martin}
D.D.E. Martin, P. Verhoeve, and A. Peacock, {\it Appl. Phys. Lett.} \textbf{88}, 123510, (2006),
DOI:10.1063/1.2187444.

\bibitem{kozorezov}
A.G. Kozorezov, and J.K. Wigmore, {\it Appl. Phys. Lett.} \textbf{89}, 223510, (2006),
DOI:10.1063/1.2397016.

\bibitem{guo}
W. Guo, et al. {\it Appl. Phys. Lett.} \textbf{110}, 212601, (2017),
DOI:10.1063/1.4984134.


\bibitem{graff}
K.F. Graff, Wave Motion in Elastic Solids, Dover, New York, (1991).


\bibitem{Hoevers}
H.F.C. Hoevers, M.L. Ridder, A. Germeau, M.P. Bruijn, P.A.J. de Korte, and R.J. Wiegerink, {\it Appl. Phys. Lett.} \textbf{86}, 251903, (2005),
DOI:10.1063/1.1949269.




\bibitem{gregory}
R.D. Gregory, {\it J. Elast.} \textbf{13}, pp. 351-355, (1983),
DOI:10.1007/BF00043002.

\bibitem{hirose}
A. Gunawan, and S. Hirose, {\it J. Acoust. Soc. Am.} \textbf{115},3,  (2004),
DOI:10.1121/1.1639330.

\bibitem{hughes}
T.J.R. Hughes, The Finite Element Method: Linear Static and Dynamic Finite Element Analysis, Dover, New York, (2000).

\bibitem{karwan1D}
K. Rostem, D.T. Chuss, K.L. Denis, and E.J. Wollack, {\it J. Phys. D. Appl. Phys.} \textbf{49}, 25, (2015),
DOI:10.1088/0022-3727/49/25/255301.

\end{thebibliography}
\end{document}